\documentclass[aps,prl,twocolumn,superscriptaddress,showpacs]{revtex4}
\usepackage{graphicx}
\usepackage{amsmath,amsfonts,amssymb}

\newcommand{\mtimes}{{$\times$}}

\newcommand{\didv}{{\textit{dI/dV}}}
\newcommand{\didvx}{{\textit{dI/dV(x)}}}

\begin{document}

\title{Strong Electron Confinement By Stacking-fault Induced Fractional Steps on Ag(111) Surfaces}
 
\author{Takashi Uchihashi}
\email{UCHIHASHI.Takashi@nims.go.jp}
\affiliation{International Center for Materials Nanoarchitectonics, National Institute for Materials Science, 1-1 Namiki, Tsukuba 305-0044, Japan}

\author{Katsuyoshi Kobayashi}
\affiliation{Department of Physics, Faculty of Science, Ochanomizu University, 2-1-1 Otsuka, Bunkyo-ku, Tokyo 112-8610, Japan}

\author{Tomonobu Nakayama}
\affiliation{International Center for Materials Nanoarchitectonics, National Institute for Materials Science, 1-1 Namiki, Tsukuba 305-0044, Japan}

\date{\today}

\begin{abstract}
The electron reflection amplitude $R$ at stacking-fault (SF) induced fractional steps is determined for Ag(111) surface states using a low temperature scanning tunneling microscope.
Unexpectedly, $R$ remains as high as $0.6 \sim 0.8$ as energy increases from 0 to 0.5 eV, which is in clear contrast to its rapidly decreasing behavior for monatomic (MA) steps [L. B{\"u}rgi et al., Phys. Rev. Lett. \textbf{81}, 5370 (1998)].
Tight-binding calculations based on {\em ab-initio} derived band structures confirm the  experimental finding. 
Furthermore, the phase shifts at descending SF steps are found to be systematically larger than counterparts for ascending steps by $\approx 0.4 \pi$.
These results indicate that the subsurface SF plane significantly contributes to the reflection of surface states.

\end{abstract}


\maketitle
The Shockley surface states on a noble metal exemplify an ideal two-dimensional (2D) electron system, which works as a basis for demonstrating and utilizing quantum natures of electrons \cite{Zangwill,Crommie_CuSW,Hasegawa_AuSW}.
To realize quantum confinement of surface states, monatomic (MA) steps \cite{Avouris_AuQSE,Li_AgIsland,Burgi_AgResonator,Mugarza_AuResonator,Morgenstern_AgSSdep}, artificially manipulated atoms \cite{Crommie_QuantumCorral,Kliewer_Resonator}, and self-assembled molecules  \cite{Pennec_SupramolecularGrating} have successfully been used.
However, the lifetimes of resultant quantized states are rather short especially at high energies, because of lossy scattering at the boundary caused by low reflection amplitude \cite{Heller_QCtheory,Pendry_1DScatterer,Burgi_AgResonator,Jensen_AgIsland,Crampin_Resonator}.
This may pose a fundamental limitation on the usage of these quantum structures.
Although the reflection at the boundary may be enhanced by multiplying potential barriers \cite{Tournier_AgNanopyramid}, a search for a new form of confinement is highly desirable.
Recently, stacking-fault (SF) defects have been found to substantially modify surface and bulk electronic states of Ag thin films \cite{Nagamura_AgStripe,Okuda_AgStripes,Sawa_AgSSDislocation,Kobayashi_SFBulk}.
Nevertheless, basic properties concerning the reflection of surface states by a SF-induced step have so far remained elusive. 

In this Letter, we determine the reflection amplitude $R$ at  SF steps for Ag(111) surface states using a low temperature scanning tunneling microscope (STM).
$R$ retains high values of $0.6 \sim 0.8$ as energy increases from 0  to 0.5 eV, which is in striking contrast to the rapid decrease in $R$ for MA steps reported in Ref. \cite{Burgi_AgResonator}.
Tight-binding calculations based on {\em ab-initio} derived band structures confirm the experimental finding.
The results demonstrate that SF steps offer a better method for realizing a strong quantum confinement on metal surfaces than MA steps.
The phase shifts at descending and ascending SF steps are also determined for the same energy region, the former being systematically larger than the latter by  $\approx 0.4 \pi$.
The possibility of significant scattering of the surface state by the subsurface SF plane is proposed based on these measurements.

\begin{figure}[b]
\includegraphics[width=75mm]{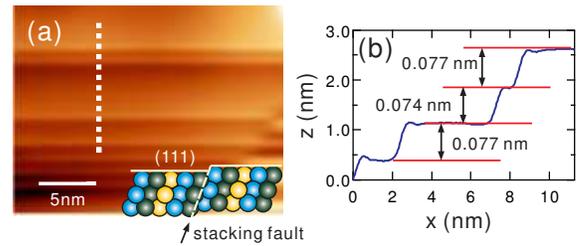}
\caption{(Color online)
(a) Typical topographic STM image of a Ag(111) film with SF induced fractional steps ($V=+1.0$V). Inset: schematic of the atomic structure of a Ag(111) film with a SF defect.
(b) Tilt-corrected height profile taken along the dashed line in (a) ($z$: height, $x$: lateral distance). 
}
\label{Fig1}
\end{figure}

The experiments were performed in an ultrahigh vacuum system equipped with a low temperature STM.
To determine the reflection amplitude and the phase shift, sufficiently long and straight SF steps are needed.
For this aim, Si(111)4\mtimes1-In surfaces (referred to as In4\mtimes1) were used as one-dimensional (1D) atomic-scale geometric templates \cite{Uchihashi_AgStripe,Nagamura_AgStripe,Okuda_AgStripes}.
Ag films about 20 monolayers (ML) in thickness were grown on In4\mtimes1 around 100 K followed by a natural annealing to room temperature.
This results in penetration of high-density SF planes into the film, which are terminated by 'fractional' steps with a height of 0.078 nm (equivalent to 1/3 of the MA step height) \cite{Uchihashi_AgStripe}. 
Figure 1 (a)(b) show a typical STM image of a Ag(111) film with SF step arrays and a tilt-corrected height profile taken along the dashed line ($z$: height, $x$: lateral distance).
Judging from their step heights, the steps are indeed of SF origin. 
In the present study, partially ordered In4\mtimes1 surfaces were used as substrates. 
This allows us to fabricate various nanostructures such as triangles and hexagons with different dimensions  and to observe surface standing waves on them.
All STM measurements were performed below 8 K.
Sample bias voltages $V$ were measured relative to the tip.
Differential conductance (\didv) spectra  were acquired by standard lock-in ac detection at a constant tip height while \didv images  were taken in the constant current mode.
Theoretical calculations of reflection amplitudes were performed using a tight-binding method, where the tight-binding parameters were determined to reproduce a band structure of a Ag(111) thin film obtained by a density-functional method.
Details of calculation methods have been described in Ref.~\cite{Kobayashi_SFBulk}.
In the present paper we used a slab of ten layers, which is sufficient to discuss the surface state.
To be compared with experimental results, reflection probability was averaged over the 1D Brillouin zone in the direction parallel to the step line.

Figure 2(a) shows a representative topographic image of a triangular island bounded by SF steps obtained at $V= -2$ V.
Topographic height $z$ measured at such a high negative voltage should closely follows the actual morphology, since the spectra of the occupied states were found to be nearly featureless.
The dashed lines indicate the position of the SF steps, which were determined from the steepest slope in topographic height \cite{Li_AgSW}. 
The straight SF lines and the three-fold symmetry imposed by the substrate lead to an ideally shaped equilateral triangle where the quantum confinement of surface states is expected \cite{Kanisawa_InAsOD}.  
The presence of quantized states in the island and their evolution with increasing energy are demonstrated by \didv images in Fig. 2(b)-(d) ((b) $V=+0.09$ V (c) $V=+0.26$ V (d) $V= +0.50$ V).
Here \didv signal was normalized to recover a quantity proportional to the surface density of states $\rho$ at a fixed height, according to the recipe prescribed by Li {\em et al.} \cite{Li_AgSW}.
Referential topographic height needed for the normalization was obtained from Fig. 2(a).
Figures 2(e)-(g) show simulated \didv\ images  based on the analytical solutions for an equilateral triangle with perfect confinement  \cite{Krishnamurthyt_Triagnle=Kumagai_AgTriangle}.
The excellent agreement between the experimental and simulated images demonstrates that such SF triangles are ideal electron resonators and can be utilized to investigate the intrinsic properties of the quantum confinement \cite{footnote_trianglephase}.

\begin{figure}
\includegraphics[width=70mm]{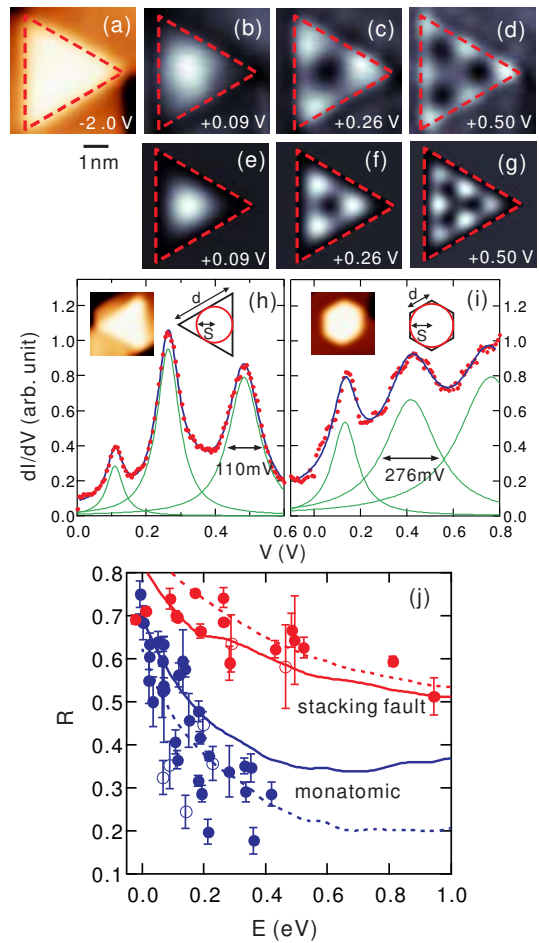}
\caption{(Color)
(a) Topographic STM image of a SF bounded equilateral triangle island. (b)-(d) Series of \didv images taken on the triangle island in (a). (e)-(g) Simulated \didv images for the same triangle island. The dashed lines indicate the positions of the steps. (h)(i) Red dots: \didv spectra taken on (h) a SF bounded equilateral triangle island and (i) a MA step bounded hexagonal island. Blue lines: fitting results using multiple Lorentzian functions. Green lines: components of the Lorentzian functions. Inset: STM images and their schematic illustrations of the islands. (j) Reflection amplitudes $R$ at SF steps (red circles) and MA steps (blue circles) as a function of energy $E$. The solid and dashed lines show tight-binding calculations for SF steps (red) and  for MA steps (blue). See the text for details.
}
\label{Fig2}
\end{figure}

Reflection amplitudes at SF steps were determined by measuring the energy widths of quantized states in triangle resonators.
This is based on the fact that electron lifetime $\tau$, relating to the energy width $\Gamma$ through an equation $\tau=\hbar/\Gamma$, is limited by incomplete reflection at the  boundary \cite{Kliewer_Resonator,Jensen_AgIsland,Crampin_Resonator}.
Figure 2(h) shows a representative \didv\  spectrum (red dots) taken on a triangle island bounded by descending SF (side length $d= 4.83$ nm).
Three sharp peaks corresponds to quantized states located at $E=0.11,0.26, 0.48$ eV.
The spectrum was fitted using multiple Lorentzian functions to obtain the full width at half maximum $\Gamma$ of each peak. 
For example, the peak width at $E=0.48$ eV was determined to be $\Gamma=110$ meV.
Similar experiments were also performed on resonators bounded by MA steps for comparison.
In this case, hexagons were chosen since they are commonly observed due to the six-hold symmetry of the Ag(111) surface. 
Figure 2(i) shows a \didv\ spectrum  (red dots) taken on a hexagonal island bounded by MA steps ($d= 1.80$ nm), which is similar in size as the triangle island. 
Although three peaks are likewise visible, they are obviously broader than those in Fig. 2(h).
The same fitting analysis using multiple Lorentzians gives $\Gamma=276$ meV for $E=0.43$ eV, being much larger than that of the SF triangle.
This indicates that the reflection amplitude is higher for SF steps than for MA steps.

Reflection amplitude $R$ was deduced from the peak width $\Gamma$ as follows.
First, a theoretical intrinsic energy width due to electron-electron and electron-phonon scattering ($5-18$ meV for $E=0.1-0.8$ eV) \cite{Eiguren_SSLifeTime=Vitali_InelasticLifeTime} was subtracted from $\Gamma$ to obtain the contribution from the boundary reflection $\Gamma_R$.
Then $R$ is calculated through the following equation:
\begin{equation}
\Gamma_R=-\frac{\hbar^2}{m^*}\sqrt{\frac{2m^* E}{\hbar^2}}\frac{\ln |R|}{S},   \label{eq:reflection_width}
\end{equation}
where $m^*=0.4m_0$ is the effective mass for Ag(111) surface states and $S$ is the radius of the largest enclosed circle within the resonator (see the inset of Figs. 2(h)(i)) \cite{Jensen_AgIsland,Crampin_Resonator,footnote_radius}.
The results are summarized in Fig. 2(j) for SF steps (red circles) and MA steps (blue circles).
The closed  and open circles represent data for descending and ascending steps, respectively. 
$R$ at MA steps rapidly decreases from about 0.7 to $0.2-0.3$ as $E$ increases from 0 to 0.4 eV,  consistent with Ref. \cite{Burgi_AgResonator}.
In clear contrast, $R$ for SF steps maintains high values around $0.6-0.8$ at least up to 0.5 eV.
Our tight-binding calculations of reflection amplitude performed for both types of steps confirmed this finding.
Red and blue lines in Fig. 2(j) show $R$ calculated for SF and MA step, respectively, the former being clearly higher than the latter (solid line: descending steps, dotted line: ascending steps).
Interestingly, the result is against an intuitive expectation that a lower step height should result in a weaker electron reflection.

To gain insight into this anomaly, we further investigated the phase shifts of electron reflection at SF steps.
The phase shift of an electron wave at a step can be obtained by analyzing the standing wave pattern formed on a terrace.
For a precise measurement, the step position was determined from the steepest slope in $z$ as described above. 
To avoid errors in $z$ and \didv at a step due to a limited feedback response, scanning was performed along the step direction and at a low speed ($\approx$ 5 nm/s).
The image was carefully monitored for a possible lateral drift and was corrected to cancel it if needed.

\begin{figure}
\includegraphics[width=60mm]{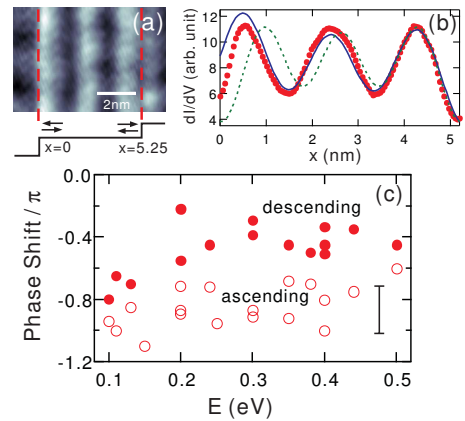}
\caption{(Color online)
(a)\didv images of surface state standing waves observed on a Ag(111) terrace caused by parallel SF steps ($V=+0.35$V). The dashed lines indicate the positions of the steps.
(b) Closed circles: normalized \didv signal deduced from the \didv image in (a). Solid and dashed lines: fitting results using the Fabry-P{\'e}rot model \cite{Burgi_AgResonator}.
(c) Phase shifts as a function of electron energy $E$ at descending and ascending SF steps with an estimated error bar. 
}
\label{Fig3}
\end{figure}

Figure 3(a) shows a \didv image of a standing wave on a terrace bounded by descending (left) and ascending (right) steps ($V=+0.35$ V), the positions of which are indicated by the dashed lines.
The \didv signal was normalized  as described above and averaged along the step direction (Fig. 3 (b), closed circles).
The graph covers the area between the two parallel steps.
The asymmetry of \didvx\ with respect to the center of the abscissa indicates different reflection behaviors at the two steps.
The phase shift at each steps were obtained using a Fabry-P{\'e}rot resonator model given by B{\"u}rgi {\em et al} \cite{Burgi_AgResonator}.
Here the ratio of bulk to surface density of states of 0.3 was experimentally determined.
To reduce the number of parameters, we let the two phase shifts $\phi_{\text{des}}$ and $\phi_{\text{asc}}$ to be independent and used an averaged reflection amplitude $R$ for both steps.
The phase shifts determined from the fitting are $\phi_{\text{des}}=-0.45\pi$ and $\phi_{\text{asc}}=-1.0\pi$ while a reflection amplitude of $R\approx 0.66$ was obtained.
Note that $R$ determined here is consistent with the previous result (Fig. 2(j)).
Figure 3(b) displays \didvx\ calculated with the above phase shifts (solid line) and that with $\phi_{\text{des}}=\phi_{\text{asc}}=-1.0\pi$ (dashed line).
The failure of the latter shows a clear deviation of $\phi_{\text{des}}$ from $-\pi$.

We repeated the same experiments while changing the electron energy $E(=eV)$.
Standing wave patterns formed by single SF steps were also analyzed to obtain phase shifts \cite{Li_AgSW,Crommie_CuSW}.
The results are summarized in Fig. 2(c). 
The phase shifts for descending steps $\phi_{\text{des}}$ are located between $-0.8 \pi \sim -0.2 \pi$, slowly increasing as $E$ increases.
These values are systematically larger than the phase shifts at ascending steps $\phi_{\text{asc}}$ by $\approx 0.4 \pi$, which are located between $-1.1 \pi \sim -0.6 \pi$.
Note that this difference cannot be explained by a shift in apparent step position due to a finite tip radius \cite{Jensen_AgIsland} because this would result in the opposite trend.
Generally, a large phase shift means that the wavefunction of electron is attracted toward the potential barrier.
This is very unlikely here because descending and ascending electrons should see the same potential barrier at SF steps, although the presence of scattering to bulk states may complicate the situation.

The large the reflection amplitude and the phase shift relation for SF steps revealed so far can be attributed to the subsurface structures of the steps.
We note that Shockley surface states on a noble metal generally extend over several atomic layers from the surface \cite{Zangwill}. 
They extend more into the bulk region as the energy increases from the band bottom and finally enter the surface resonance region.
Considering that a MA step is a surface-localized defect and retains the bulk periodicity below the surface, its reflection amplitude will descrease rapidly in this energy region  \cite{footnote_EffectiveBarrier}.
In contrast, a SF studied here, created by the substrate template, should penetrate to the bottom of the film.
The subsurface defect of a SF step may strongly reflect surface states even if they extend deep into the bulk region at high energies.
This is plausible because strong anisotropic modulation of {\em bulk} electronic states of Ag films by periodic SF planes has been found experimentally \cite{Nagamura_AgStripe,Okuda_AgStripes} and  theoretically \cite{Kobayashi_SFBulk}.
The relation between the phase shifts at descending and ascending steps, $\phi_{\text{des}}>\phi_{\text{asc}}$, supports such an interpretation.
The subsurface part of the SF plane extends towards the lower side of the step (see the inset of Fig. 1 (a)).
If surface states are reflected strongly by the subsurface SF, the effective potential barrier should be shifted in the descending direction, which will result in the observed relation $\phi_{\text{des}}>\phi_{\text{asc}}$.

In summary, we have demonstrated  high reflection amplitudes at SF steps around $0.6 \sim 0.8$ for $E=0 \sim 0.5$ eV, unexpectedly higher than those for MA steps.
The relation between the phase shifts at descending and ascending SF steps $\phi_{\text{des}}>\phi_{\text{asc}}$ was also clarified, which points to the significant scattering by the subsurface SF plane.
 The present result offers a promising route for effective electron confinement and fabrication of various quantum structures on metal surfaces.
By increasing the size of a resonator and thereby reducing the effect of the lossy scattering, for example, the intrinsic lifetime of electrons could be determined more accurately \cite{Crampin_Resonator,Tournier_AgNanopyramid}.

We thank R. Berndt, J. Kr{\"o}ger, and O. Groening for fruitful discussions and M. Becker for his technical support in the early stage.
T. U. acknowledges JSPS for the financial support KAKENHI Grant No. 21510110 and Iketani Science and Technology Foundation.


\begin{references}
\bibitem{Zangwill}
	A. Zangwill, {\em Physics at Surfaces}, Chapt. 4 (Cambridge Univ. Press, 1988).

\bibitem{Crommie_CuSW}
	M. F. Crommie, C. P. Lutz, and D. M. Eigler,
	Nature 	\textbf{363}, 524 (1993).

\bibitem{Hasegawa_AuSW}
	Y. Hasegawa and Ph. Avouris,
	Phys. Rev. Lett. \textbf{71}, 1071 (1993).


\bibitem{Avouris_AuQSE}
	Ph. Avouris and I.-W. Lyo,
	Science	\textbf{264}, 942 (1994).

\bibitem{Li_AgIsland}
	J. Li, W. D. Schneider, R. Berndt, and S. Crampin,
	Phys. Rev. Lett. \textbf{80}, 3332 (1998).


\bibitem{Burgi_AgResonator}
	L. B{\"u}rgi, O. Jeandupeux, A. Hirstein, H. Brune, and K. Kern,
	Phys. Rev. Lett. \textbf{81}, 5370 (1998).

\bibitem{Mugarza_AuResonator}
	A. Mugarza, A. Mascaraque, V. P{\'e}rez-Dieste, V. Repain, S. Rousset, F. J. Garc{\'i}a de Abajo, and J. E. Ortega
	Phys. Rev. Lett. \textbf{87}, 107601	(2001);
A. Mugarza and J. E. Ortega, 
	J. Phys.: Condens. Matter \textbf{15} (2003) S3281.

\bibitem{Morgenstern_AgSSdep}
K. Morgenstern, K.-F. Braun, and K.-H. Rieder,
	Phys. Rev. Lett. \textbf{89}, 226801 (2002).

\bibitem{Crommie_QuantumCorral}
	M. F. Crommie, C. P. Lutz, and D. M. Eigler,
	Science	\textbf{262}, 218 (1993).
	

\bibitem{Kliewer_Resonator}
	J. Kliewer, R. Berndt, and S. Crampin,
	New. J. Phys. \textbf{3}, 22 (2001).


\bibitem{Pennec_SupramolecularGrating}
Y. Pennec, W. Auw{\"a}rter, A. Schiffrin, A. Weber-Bargioni, A. Riemann, and J. V. Barth,
Nature Nanotech. \textbf{2} 99 (2007). 

\bibitem{Heller_QCtheory}
	E. J. Heller, M. F. Crommie, C. P. Lutz, and D. M. Eigler,
	Nature	\textbf{369}, 464 (1994).

\bibitem{Pendry_1DScatterer}
	G. H{\"o}rmandinger and J. B. Pendry,
	Phys. Rev. B \textbf{50}, 18607 (1994).

\bibitem{Jensen_AgIsland}
	H. Jensen, J. Kr{\"o}ger, R. Berndt, and S. Crampin,
	Phys. Rev. B \textbf{71}, 155417 (2005).

\bibitem{Crampin_Resonator}
	S. Crampin, H. Jensen, J. Kr{\"o}ger, L. Limot, and R. Berndt,
	Phys. Rev. B \textbf{72}, 035443 (2005).

\bibitem{Tournier_AgNanopyramid}
C. Tournier-Colletta, B. Kierren, Y. Fagot-Revurat, and D. Malterre,
Phys. Rev. Lett. \textbf{104}, 016802 (2010).


\bibitem{Nagamura_AgStripe}
	N. Nagamura, I. Matsuda, N. Miyata, T. Hirahara, S. Hasegawa, and T. Uchihashi,
	Phys. Rev. Lett. \textbf{96}, 256801 (2006).

\bibitem{Okuda_AgStripes}
T. Okuda, Y. Takeichi, K. He, A. Harasawa, A. Kakizaki, and I. Matsuda,
Phys. Rev. B \textbf{80}, 113409 (2009).

\bibitem{Sawa_AgSSDislocation}
	K. Sawa, Y. Aoki, and H. Hirayama,
	Phys. Rev. Lett. \textbf{104}, 016806 (2010).

\bibitem{Kobayashi_SFBulk}
K. Kobayashi and T. Uchihashi,
	Phys. Rev. B \textbf{81}, 155418 (2010).

\bibitem{Uchihashi_AgStripe}
	T. Uchihashi, C. Ohbuchi, S. Tsukamoto, and T. Nakayama,
	Phys. Rev. Lett. \textbf{96}, 136104 (2006).


\bibitem{Li_AgSW}
	J. Li, W. D. Schneider, and R. Berndt,
	Phys. Rev. B \textbf{56}, 7656 (1997).




\bibitem{Kanisawa_InAsOD}
K. Kanisawa, M. J. Butcher, Y. Tokura, H. Yamaguchi, and Y. Hirayama,
	Phys. Rev. Lett. \textbf{87}, 196804 (2001). 


\bibitem{Krishnamurthyt_Triagnle=Kumagai_AgTriangle}
H. R. Krishnamurthyt, H. S. Mani, and H. C. Verma,
J. Phys. A Math. Gen. \textbf{15}, 2131 (1982);
T. Kumagai and A. Tamura,
J. Phys. Soc. Jpn. \textbf{77}, 014601 (2008). 


\bibitem{footnote_trianglephase}
The larger sizes of the triangular patterns for experiments compared to those for simulations are due to the significant deviation from $-\pi$ for the phase shift at descending SF steps, as described later.


\bibitem{Eiguren_SSLifeTime=Vitali_InelasticLifeTime}
A. Eiguren, B. Hellsing, F. Reinert, G. Nicolay, E.V. Chulkov, V. M. Silkin, S. H{\"u}fner, and P. M. Echenique;
Phys. Rev. Lett. \textbf{88}, 066805 (2002).
L. Vitali, P. Wahl, M. A. Schneider, K. Kern, V. M. Silkin, E.V. Chulkov, P. M. Echenique, Surf. Sci. \textbf{523} L47 (2003).


\bibitem{footnote_radius}
The radius $S$ for a SF triangle island was determined from quantized energy levels rather than from its geometrical size. A choice of the geometrical size would result in even higher reflection amplitudes, thus not affecting the conclusion. 



\bibitem{footnote_EffectiveBarrier}
A similar argument was given in Ref.\cite{Mugarza_AuResonator} concerning the strength of the effective potential barrier of MA steps.

\end{references}
\end{document}